\pdfoutput=1
\RequirePackage{lineno}
\newif\ifpreprint%
\preprintfalse%
\ifpreprint%
	\documentclass[preprint,english,aps,prb,floatfix,amssymb,linenumbers,superscriptaddress]{revtex4}
\else%
	\documentclass[twocolumn,english,aps,prb,floatfix,amssymb,linenumbers,superscriptaddress]{revtex4}
\fi%
\usepackage{amsmath}
\usepackage{dsfont}
\usepackage{color}
\usepackage{amssymb}
\usepackage{graphicx}
\usepackage{braket}
\usepackage{units}
\ifpreprint%
\else%
\fi%

\newcommand{\centerFrequency}{73.5 }
\newcommand{\lowerGap}{72.92 }
\newcommand{\upperGap}{74.89 }
\newcommand{\arbitraryFrequency}{72.0 }
\newcommand{\site}{10 }
\newcommand{\siteL}{5}
\newcommand{\Wthickness}{0.364 }
\newcommand{\WthicknessMu}{364 }
\newcommand{\plateOffset}{73.895\pm0.03}
\newcommand{\weakHop}{1.89\pm0.07\times10^9}
\newcommand{\strongHop}{6.69\pm0.17\times10^9}
\newcommand{\paramPlateSize}{5 mm x 5 mm x 0.364 mm }

\ifx\pdftexversion\undefined%
\usepackage[dvips]{hyperref}
\else%
\usepackage{hyperref}
\fi%
\hypersetup{
  colorlinks = false,
  hypertexnames=false
}

\renewcommand{\theta}{\vartheta}
\renewcommand{\phi}{\varphi}

\begin{document}
\ifpreprint%
	\linenumbers%
\fi%

\title{Observation of a phononic quadrupole topological insulator}

\date{\today}

\author{Marc Serra-Garcia}
\thanks{These authors contributed equally to this work.}
\affiliation{Institute for Theoretical Physics, ETH Zurich, 8093 Z\"urich, Switzerland}
\author{Valerio Peri}
\thanks{These authors contributed equally to this work.}
\affiliation{Institute for Theoretical Physics, ETH Zurich, 8093 Z\"urich, Switzerland}
\author{Roman S\"usstrunk}
\affiliation{Institute for Theoretical Physics, ETH Zurich, 8093 Z\"urich, Switzerland}
\author{Osama R. Bilal}
\affiliation{Institute for Theoretical Physics, ETH Zurich, 8093 Z\"urich, Switzerland}
\affiliation{Division of Engineering and Applied Science, California Institute of Technology, Pasadena, CA 91125, USA}
\author{Tom Larsen}
\affiliation{Advanced NEMS Group, \'Ecole Polytechnique F\'ed\'erale de Lausanne (EPFL), 1015 Lausanne, Switzerland}
\author{Luis Guillermo Villanueva}
\affiliation{Advanced NEMS Group, \'Ecole Polytechnique F\'ed\'erale de Lausanne (EPFL), 1015 Lausanne, Switzerland}
\author{Sebastian D. Huber}
\affiliation{Institute for Theoretical Physics, ETH Zurich, 8093 Z\"urich, Switzerland}


\maketitle

{\bf 
The modern theory of charge polarization in solids \cite{King-Smith93, Taherinejad14} is based on a generalization of Berry's phase.\cite{Berry84} Its possible quantization\cite{Thouless82,Kane05a} lies at the heart of our understanding of all systems with topological band structures that were discovered over the last decades.\cite{Chiu16, Klitzing80, Konig07, Hsieh08, Xu15} While based on the concept of the ``charge'' polarization, the same theory can be used as an elegant tool to characterize the Bloch bands of neutral bosonic systems such as photonic\cite{Lu16} or phononic crystals.\cite{Susstrunk15, Susstrunk16} Recently, the theory of this quantized polarization was extended from the dipole- to higher multipole-moments.\cite{Benalcazar17} In particular, a two-dimensional quantized quadrupole insulator is predicted to have gapped yet topological one-dimensional edge-modes, which in turn stabilize zero-dimensional in-gap corner states.\cite{Benalcazar17} However, such a state of matter has not been observed experimentally. Here, we provide the first measurements of a phononic quadrupole insulator. We experimentally characterize the bulk, edge, and corner physics of a mechanical metamaterial and find the predicted gapped edge and in-gap corner states. We further corroborate our findings by comparing the mechanical properties of a topologically non-trivial system to samples in other phases predicted by the quadrupole theory. From an application point of view, these topological corner states are an important stepping stone on the way to topologically protected wave-guides\cite{Susstrunk15, Nash15} in higher dimensions and thereby open a new design path for metamaterials.\cite{Liu00, Huber16}
}
%
%
\ifpreprint%
\else%
\begin{figure}[b!]
\includegraphics{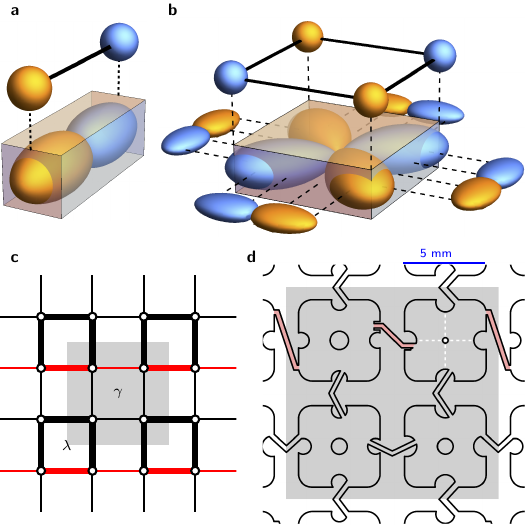}
\input{fig1caption}
\label{fig:setup}
\end{figure}
\fi%

A non-vanishing dipole moment ${\bf p}=\langle \Psi|{\bf r}|\Psi \rangle$ in an insulator does not lead to any charge accumulation in the bulk. However, it manifests itself through uncompensated surface charges and hence induces potentially interesting surface physics, see Fig.~\ref{fig:setup}a. The dipole moment ${\bf p}$ is expressible through Berry's phase,\cite{King-Smith93,Berry84} which in turn can lead to its quantization.\cite{Thouless82,Kane05a,Bernevig06a,Fu06,Moore07,Qi08} All observed topological insulators fit into this framework of quantized dipole moments,\cite{Thouless82} or mathematical generalizations thereof.\cite{Qi08} Moreover, for neutral systems, the abstract quantity ${\bf p}$ loses its electromagnetic content. However, it can equally well be used to predict band-structure effects such as stable surface modes. Whether higher order moments, such as the quadrupole, can lead to distinctly new topological phases of matter remained unclear.
\ifpreprint%
\else%
\begin{figure*}[t!]
\includegraphics{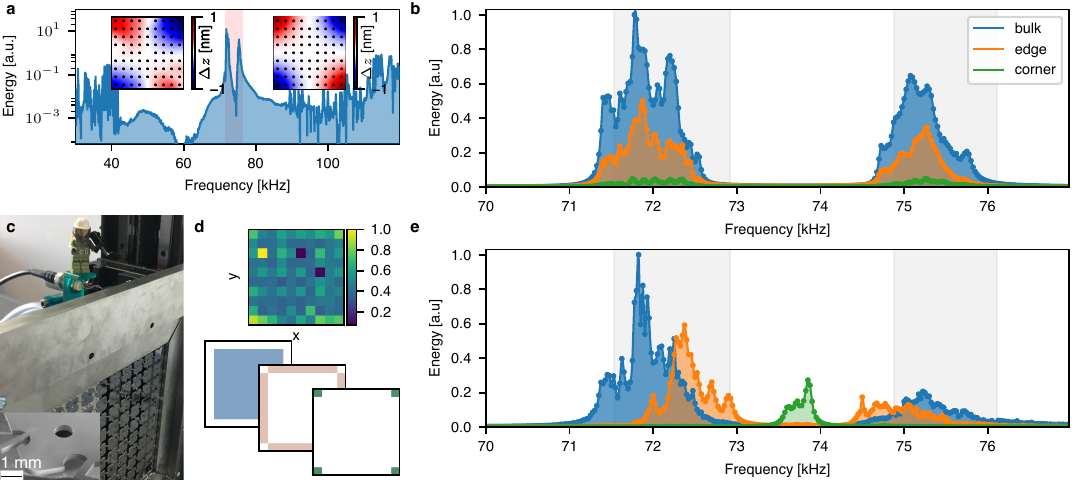}
\input{fig2caption}
\label{fig:midgap}
\end{figure*}
\fi%

Recently, a theory for a quantized quadrupole insulator was put forward\cite{Benalcazar17} based on its phenomenology: A bulk quadrupole moment in a finite two-dimensional sample gives rise to surface dipole moments on its one-dimensional edges as well as to uncompensated charges on the zero-dimensional corners, see Fig.~\ref{fig:setup}b. The former is indicating {\em gapped edge modes} while the latter motivates the presence of {\em in-gap corner excitations}. This also defines the key technological use of such a quadrupole insulator in mechanical or optical metamaterials: The localized corner modes can be used for acoustic field enhancement in two dimensions.\cite{Xiao15} Moreover, these states serve as a stepping stone towards topologically protected, one-dimensional channels in three dimensions: When appropriately stacked into three dimensions, the corner modes give rise to chiral one dimensional modes along edges of the three dimensional sample.\cite{Huber16, Rechtsman13, Hafezi13, Lang17, Benalcazar17a}

The phenomenology of gapped edges and gapless corners can be formalized mathematically. Benalcazar et al.\cite{Benalcazar17} proposed to use nested Wilson loops as a way to obtain a quantized quadrupole moment (see Methods for details): Wilson loop operators depend only on the bulk properties and encode the edge physics via their eigenvalues $\nu^\pm(k_\alpha)$, $\alpha=x,y$, known as Wannier bands.\cite{Fidkowski11} If these Wannier bands $\nu^\pm(k_\alpha)$ are gapped, the eigenvectors of the Wilson loops can be used to define the bulk-induced edge polarization of the bands below the gap $p_\alpha^{\nu_-}$, where $\alpha$ denotes the direction of the polarization. In the same way as for the conventional topological insulators,\cite{Kane05a} symmetries are required for the quantization of $p_\alpha^{\nu_-}$. Particularly, the presence of inversion symmetry $I$ and non-commuting mirror symmetries $M_x$ and $M_y$ lead to a well defined and quantized $p_\alpha^{\nu_-} \in \{ 0,1/2 \}$. In particular, the sought after quantized quadrupole phase is described by\cite{Benalcazar17}
\begin{linenomath}
\begin{equation}
(p_x^{\nu_-},p_y^{\nu_-}) = (1/2,1/2).   
\end{equation}
\end{linenomath}
As a corner terminates two edges, $\left(p_x^{\nu_-},p_y^{\nu_-}\right) = \left(1/2,1/2\right)$ could suggest that each of them supports two in-gap states. However, it is an important hall-mark of the bulk nature of the quadrupole insulator that each corner hosts only one mode, cf. Fig.~\ref{fig:setup}b.\cite{Benalcazar17}
\ifpreprint%
\else%
\begin{figure}[t!]
\includegraphics{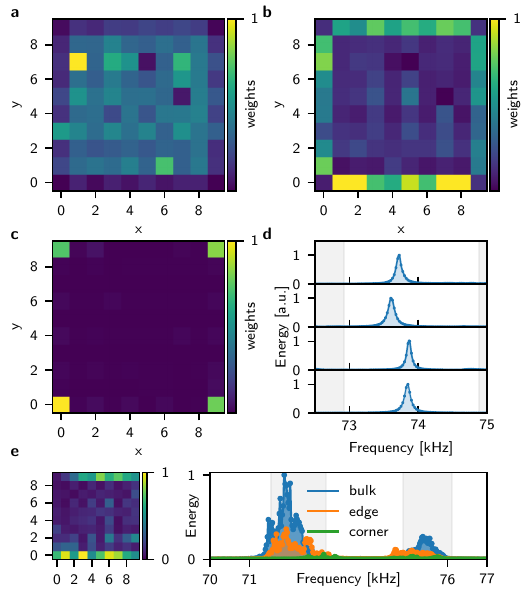}
\input{fig3caption}
\label{fig:modes}
\end{figure}
\fi%

A concrete tight-binding model for a two-dimensional quantized quadrupole insulator is shown in Fig.~\ref{fig:setup}c.\cite{Benalcazar17} The dimerized hopping with amplitude $\lambda$ and $\gamma$ leads to a band-gap between two pairs of degenerate bands for $\lambda \neq \gamma$ (see Methods). The black (red) lines in Fig.~\ref{fig:setup}c indicate positive (negative) hoppings, effectively emulating  a magnetic $\pi$-flux per plaquette. This $\pi$-flux requires the mirror-symmetry around the horizontal axis ($M_y$) to be accompanied by a gauge-transformation, leading to the non-commutation of $M_x$ and $M_y$. The present model also has inversion $I$ and $C_4$ rotational symmetry (again up to a gauge-transformation) forcing $p_x^{\nu_-} = p_y^{\nu_-}$ as well as particle-hole symmetry fixing the corner modes to the middle of the gap. For $\gamma<\lambda$ the topological phase $(p_x^{\nu_-},p_y^{\nu_-})=(1/2,1/2)$, whereas for $\gamma>\lambda$, the trivial phase $(0,0)$ is realized.\cite{Benalcazar17} Here, we seek a mechanical implementation of a quadrupole insulator with $\ddot x_i=-\mathcal D_{ij}  x_j$, where the dynamical matrix $\mathcal D_{ij}$ couples local degrees of freedom $x_i$ according to the model in Fig.~\ref{fig:setup}c.

We implement the quadrupole insulator using the concept of perturbative mechanical metamaterials.\cite{Matlack16} The starting point is a single-crystal silicon plate with dimensions \siteL$\times$\siteL$\times$\Wthickness\,mm, whose mechanical eigenmodes are described by the displacement field ${\bf u}({\bf r})$. We work with the first non-rigid-body mode which is characterized by two perpendicular nodal lines in the out-of-plane component of ${\bf u}({\bf r})$, see Fig.~\ref{fig:setup}d and Fig.~\ref{fig:midgap}a. By spectrally separating this mode from the modes below and above it, one can describe the dynamics in some frequency range by specifying only the amplitude $x_i$ of the mode of interest of a given plate $i$. The hopping elements in $\mathcal D_{ij}$ are then implemented by thin beams between neighboring plates. The nodal structure of the mode allows to mediate couplings of either positive or negative sign, depending on which sides of the nodal lines are connected by the beams. Moreover, the distance to the nodal line controls the coupling strength mediated by a given beam. Combinatorial search\cite{Coulais16} followed by a gradient optimization\cite{Matlack16} leads to the design in Fig.~\ref{fig:setup}d which is characterized by a ratio $|\gamma/\lambda|=0.28$, or $|\lambda/\gamma|=0.28$, see Methods.
 
All measurements shown are performed using the same scheme: The plates are excited with an ultrasound air-transducer. The transducer has a diameter of 5\,mm and is in close proximity to the sample, such that only a single plate is excited. We measure the response of the excited plate with a laser-interferometer. In this way, we measure the out-of-plane vibration amplitude $\Delta z_i \propto \psi_i^2$, where $\psi_i$ is the eigenmode at the measured frequency (both the excitation strength and the measurement scale with $\psi_i$). The inset of Fig.~\ref{fig:midgap}a shows the local mode of a single plate measured in this way. In all other figures we display the mechanical energy $\varepsilon_i\propto \Delta z_i^2$.

To identify the in-gap states we take a measurement of $\varepsilon_i(\nu)$ as a function of frequency $\nu$ on all plates $i$. We then apply the filters $\varepsilon_\alpha(\nu)=\sum_i \varepsilon_i(\nu)F_{i,\alpha}$ shown in Fig.~\ref{fig:midgap}d to separate the response of the bulk, edges, and corners. Figs.~\ref{fig:midgap}b \& e show the resulting spectra for two different samples (see Methods). In the topologically trivial case with $\gamma > \lambda$, one can observe two frequency bands where the system absorbs energy (the theoretically predicted location of the bands is indicated in gray). Two features characterize this trivial phase: (i) No frequency range is dominated by the edge or corner response. Moreover, the relative weight of the three curves is in good accordance with the respective number of sites in the bulk, edges, and corners, respectively. (ii) No resonances appear in the gap between \lowerGap\,kHz and \upperGap\,kHz. For the sample with $\gamma<\lambda$ in Fig.~\ref{fig:midgap}e, two key-features of the quantized quadrupole phase appear: (ii) close to \lowerGap\,kHz and \upperGap\,kHz, the response is dominated by the edges, indicative of the bulk-induced gapped edge modes. (ii) Sharp resonances at the corners appear in the gap region. A small mirror symmetry breaking leads to the non-degeneracy of the in-gap states which we discuss below.
  
The spectra in Fig.~\ref{fig:midgap}b \&\ e allow to identify three frequency regions $\mathcal B$, $\mathcal E$, and $\mathcal C$, where the bulk (blue), edge (orange), or corner (green) response dominates. To establish the quadrupole nature of the metamaterial, we analyze the site-dependent, frequency integrated response $\varepsilon_i^\alpha=\sum_{\nu\in\alpha} \varepsilon_i(\nu)$ with $\alpha=\mathcal B,\mathcal E,\mathcal C$. In Fig.~\ref{fig:modes}a--c we show the resulting spatial profiles. Note that the bulk induces gapped edge-modes on all four sides of the sample. 

The hallmark of the quadrupole phase lies in the counting of corner modes: Each corner terminates two-gapped edges, nevertheless, they all host only one in-gap mode.\cite{Benalcazar17} In Fig.~\ref{fig:modes}d, we show the response $\varepsilon(\nu)$ for the four corner plates. The resonances in the four corners are split by the presence of next-to-nearest neighbor couplings that break the particle-hole symmetry. However, each corner hosts only one resonance peak. Moreover, measurements of the edge-Greens function further support this claim, see Methods. 

To corroborate our claim of observing a quadrupole insulator, we further explore the phase diagram of Ref. \onlinecite{Benalcazar17}. When the $C_4$-symmetry is broken by allowing for different hoppings in $x$- and $y$-direction (see Methods), the phase $(p_x^\nu,p_y^\nu)=(1/2,0)$ can be reached via a gap-closing of the surface modes. The $(1/2,0)$-phase is characterized by gapped edge spectra on two parallel edges and no emergent edge physics on the perpendicular surfaces.\cite{Benalcazar17} Moreover, the induced edge modes are in a trivial state and no corner charges are induced. In Fig.~\ref{fig:modes}e, we show measurements on a sample in the $(1/2,0)$-phase, where no in-gap states appear and the frequency region dominated by the edges draws its weight from only two surfaces.

\ifpreprint%
\else%
\begin{figure}[tb]
\includegraphics{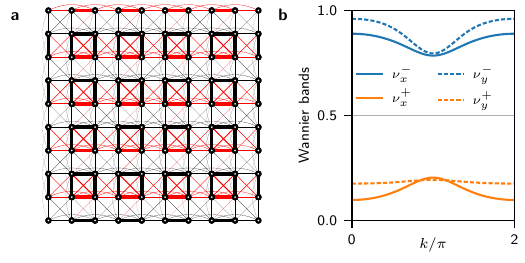}
\input{fig4caption}
\label{fig:wannier}
\end{figure}
\fi%
In addition to the experimental data presented above, we also validate our system through extensive numerical calculations. The design process for the sample shown in Fig.~\ref{fig:setup}d requires a finite-element simulation of the displacement fields ${\bf u}_i({\bf r})$ on four unit cells containing a total of 16 sites $i$. The modes obtained in this way can then be projected onto the basis of uncoupled plate-modes ${\bf u}_{i}^0({\bf r})$. In this way a reduced order model $\tilde{\mathcal D}_{ij}$ in the frequency range of the modes ${\bf u}_{i}^0({\bf r})$ is obtained.\cite{Matlack16} In Fig.~\ref{fig:wannier}a, we show the resulting model extended to a \site\!\!$\times$\site system. The nearest neighbor couplings indeed follow the blueprint of the target model shown in Fig.~\ref{fig:setup}c. However, spurious long-range couplings mediated by off-resonant admixing of other single-plate modes induce a certain amount of mirror-symmetry breaking. This is most notable in the $y$-direction, where negative next-to-nearest neighbor couplings are mapped to positive ones, which is not corrected for in the gauge-transformation in $M_y$.

The reduced order model $\tilde{\mathcal D}_{ij}$ can also be used to calculate the topological indices $(p_x^\nu,p_y^\nu)$. The gapped Wannier bands $\nu_x^\pm(k_y)$ and $\nu_y^\pm(k_x)$ are shown in Fig.~\ref{fig:wannier}b. Note that the $M_x$ symmetry implies $\nu_x^+(k_y)+\nu_x^-(k_y)=1/2$ and the same for $x\leftrightarrow y$.\cite{Benalcazar17} The absence of an exact $M_y$ symmetry indeed leads to a breaking of this rule. This is also reflected in the value of the polarizations
\begin{linenomath}
\begin{equation}
        (p_x^{\nu_-},p_y^{\nu_-}) = (0.50,0.56).
\end{equation}
\end{linenomath}
As expected from the structure of $\tilde{\mathcal D}_{ij}$ shown in Fig.~\ref{fig:wannier}a, the polarizations are not precisely quantized. However, the phenomenology of in-gap corner modes is still observed as the symmetry breaking terms do not lead to any gap-closing, neither on the edge nor in the bulk.

The results presented in this paper underline the power of perturbative metamaterials.\cite{Matlack16} On one hand, we leveraged this technique to find a first implementation of a quantized quadrupole insulator, a new class of topological materials. On the other hand, the platform of a continuous elastic medium provides a direct route to technological applications for any theoretical idea which can be represented by a tight-binding model. 


\bigskip
\noindent
{\bf Acknowledgements} We acknowledge financial support from the Swiss National Science Foundation and the NCCR QSIT. TL acknowledges support from a Marie Curie fellowship and ORB the ETH postdoctoral fellowship FEL-26 15-2.

\smallskip
\noindent
{\bf Author contributions} SDH conceived the research. MSG, VP and ORB designed the samples. MSG, VP, SDH, and SR conducted the experiments. LGV and TL fabricated the samples.

\bibliographystyle{naturemag}
\bibliography{ref}

\ifpreprint%
\begin{widetext}
\newpage
\begin{figure}[t]
\input{fig1caption}
\label{fig:setup}
\end{figure}
\begin{figure}[t]
\input{fig2caption}
\label{fig:midgap}
\end{figure}
\begin{figure}[t]
\input{fig3caption}
\label{fig:modes}
\end{figure}
\begin{figure}[t]
\input{fig4caption}
\label{fig:wannier}
\end{figure}
\end{widetext}
\fi%
\renewcommand{\figurename}{Extended data Fig.} \setcounter{figure}{0}%
\ifpreprint%
\begin{widetext}
\newpage
\begin{figure}[t]
\input{extendedfig1caption}
\label{fig:ExtendedTheory}
\end{figure}
\begin{figure}[t]
\input{extendedfig2caption}
\label{fig:ExtendedTransducer}
\end{figure}
\begin{figure}[t]
\input{extendedfig3caption}
\label{fig:ExtendedG}
\end{figure}
\begin{figure}[t]
\input{extendedfig4caption}
\label{fig:ExtendedGE}
\end{figure}
\begin{figure}[t]
\input{extendedfig5caption}
\label{fig:ExtendedDesign}
\end{figure}
\end{widetext}
\fi%
\clearpage

\medskip
\noindent
{\bf \large Methods}

\ifpreprint%
\else%
\begin{figure*}[t!]
\includegraphics{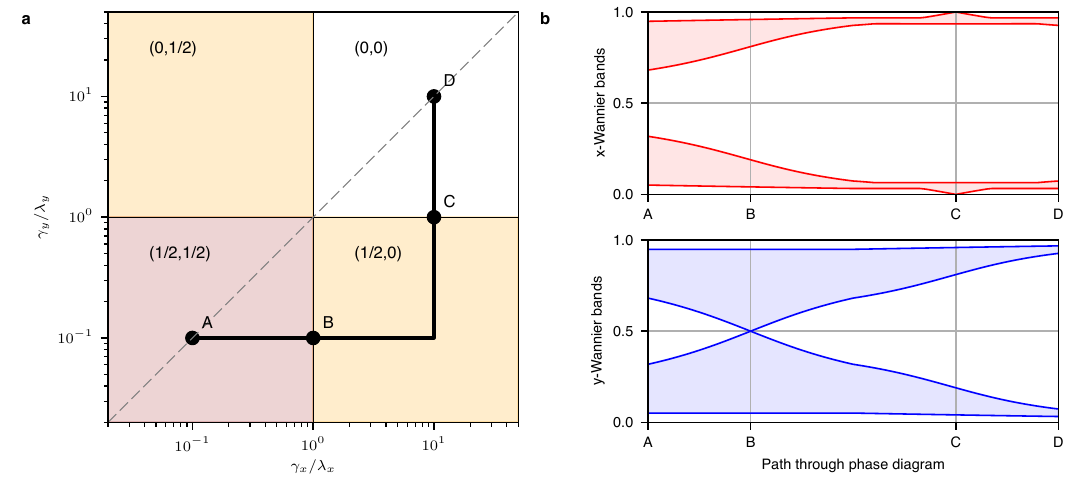}
\input{extendedfig1caption}
\label{fig:ExtendedTheory}
\end{figure*}
\fi%
\noindent
{\bf Topological quantum number: Nested Wilson loops.} Here we use the language of fermions, where bands below a given gap can be ``filled''. For the phononic case, one has to replace ``filled bands'' with ``bands below the frequency of interest''. Assuming two bands $n=1,2$ are filled, one can use the non-abelian Berry phase $\mathcal A^x_{nm}({\bf k})=i\langle u_{m}({\bf k})|\partial_{k_{x}}|u_{n}({\bf k})\rangle$ of the Bloch wave-functions $|u_{n}({\bf k})\rangle$ to construct the Wilson-loop operators
\begin{linenomath}\begin{equation}
 \label{eqn:wilson}
 \mathcal W_x(k_y) = {\rm T} \exp\left[i\oint dk_x\, \mathcal A^x_{nm}({\bf k})\right].   
\end{equation}\end{linenomath}
Here, ${\rm T}$ denotes the path ordering along a closed loop in the Brillouin zone. The eigenvalues $\nu^{\pm}(k_y)$ of $\mathcal W_x(k_y)$ are in one-to-one correspondence to the spectrum of an edge perpendicular to the $x$-coordinate\cite{Fidkowski11} (or perpendicular to $y$ when $x$ and $y$ are interchanged). If the edge modes are gapped, the eigenvectors $v^{\pm}_{n}(k_y)$ of $\mathcal W_x(k_y)$ can be used to split the filled bands in a well-defined way: $|w_{\pm}({\bf k})\rangle=\sum_{n=1}^2 v^{\pm}_n(k_y)|u_{n}({\bf k})\rangle$. The nested polarization is then defined as
\begin{linenomath}\begin{equation}
    p_y^{\nu_\pm}  = \frac{1}{(2\pi)^2} \int d{\bf k}\,  \mathcal A_\pm^y({\bf k}),
\end{equation}\end{linenomath}
with $\mathcal A_\pm^y({\bf k})=i\langle w_\pm({\bf k})|\partial_{k_{y}}| w_\pm({\bf k})\rangle$. It can be shown that the presences of two mirror-symmetries $M_x$ and $M_y$ that do not commute are a necessary requirement for the nested polarizations $p_x^{\nu_\pm}$ and $p_y^{\nu_\pm}$ to be quantized to 0 or $1/2$.\cite{Benalcazar17} 

\noindent
{\bf Model.} The model shown in Fig.~\ref{fig:setup}c can be expressed with the help of $\Gamma$-matrices $\Gamma_k=-\tau_2 \otimes \sigma_k$, $\Gamma_4=\tau_1 \otimes \sigma_0$, $k=1,2,3$; $\tau,\sigma$ are the standard Pauli-matrices. Using these matrices we can write\cite{Benalcazar17}
\begin{linenomath}
\begin{multline}
    \label{eqn:model}
    \mathcal D(k_x,k_y) = [\gamma_x + \lambda_x \cos(k_x)]\Gamma_4 + \lambda_x \sin(k_x)\Gamma_3 \\
                +[\gamma_y + \lambda_y \cos(k_y)]\Gamma_2 + \lambda_y \sin(k_y)\Gamma_1 
               =\sum_{i=1}^4 d_i({\bf k})\Gamma_i.
\end{multline}
\end{linenomath}
The $C_4$-symmetric version of Fig.~\ref{fig:setup}c is obtained by setting $\lambda_x=\lambda_y$ and $\gamma_x=\gamma_y$. The mirror symmetries are represented by $\mathcal D(-k_x,k_y)=m_x \mathcal D(k_x,k_y)m_x^\dag$ and $\mathcal D(k_x,-k_y)=m_y \mathcal D(k_x,k_y)m_y^\dag$ with $m_x=\tau_1\sigma_3$ and  $m_y=\tau_1\sigma_1$, respectively. The eigenvalues of $\mathcal D(k_x,k_y)$ are given by $\zeta=\pm |{\bf d}({\bf k})|$, leading to two doubly-degenerate bands. Bulk gap-closings occur when ${\bf d}({\bf k})=0$, which only happens for the $C_4$-symmetric case at $\lambda=\pm \gamma$. The spectrum of the mechanical system is given by $\nu=\sqrt{\nu_0^2+\zeta}$, with a frequency offset $\nu_0$. Finally, the eigenvectors $|u_n({\bf k})\rangle$ of $\mathcal D(k_x,k_y)$ can be used to calculate the Wilson loop operators of Eq. (\ref{eqn:wilson}). The phase diagram and the evolution of the Wannier bands of model (\ref{eqn:model}) are shown in the Extended Data Fig.~\ref{fig:ExtendedTheory}.

The decay of the edge and corner states into the bulk is simple to derive in analogy to the Su-Schrieffer-Heeger model,\cite{Su79} where the wavefunction has a node on every other site and is exponentially decaying with a decay length of $\xi/a = 2/(\log|\lambda/\gamma|)$ (in units of the site-to-site distance $a$). For our ratio of of $\lambda/\gamma\approx 0.28$ (see below) this results in $\xi/a\approx 1.6$. In other words, $1-e^{-4/\xi}\approx 92\%$ of the energy of a edge/corner mode is stored on the outermost row/corner site.

\ifpreprint%
\else%
\begin{figure*}[tbh]
\includegraphics{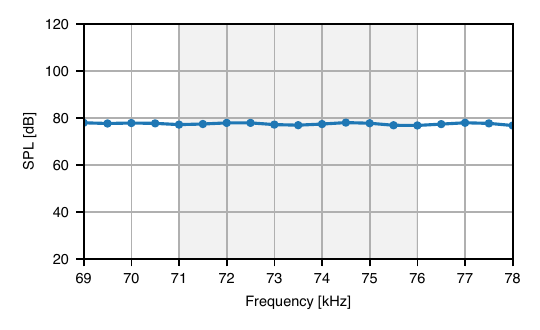}
\input{extendedfig2caption}
\label{fig:ExtendedTransducer}
\end{figure*}
\fi%
\ifpreprint%
\else%
\begin{figure*}[tbh]
\includegraphics{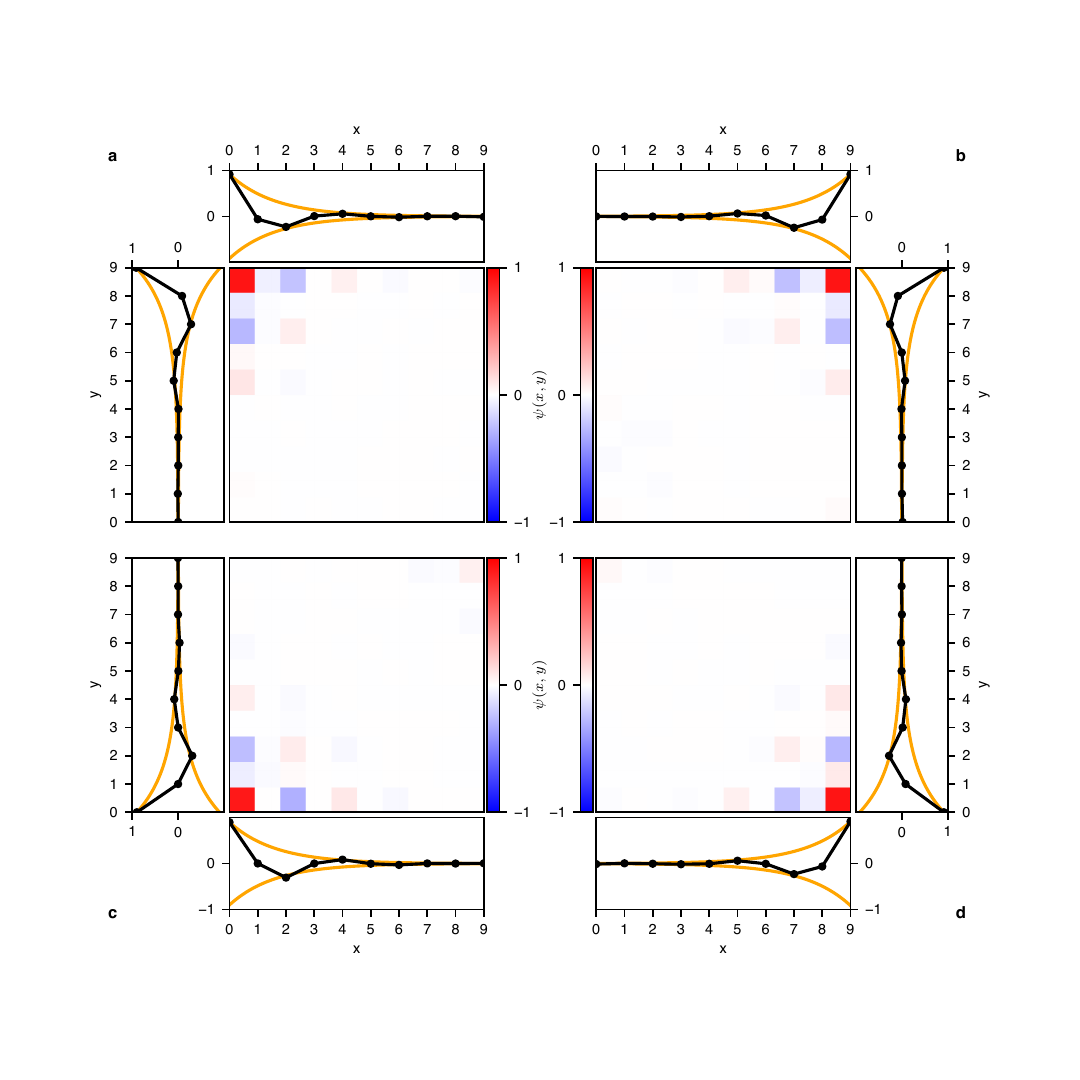}
\input{extendedfig3caption}
\label{fig:ExtendedG}
\end{figure*}
\fi%
\noindent
{\bf Signal analysis.} All measurements are performed with an interferometer (IDS3010 from attocube) after exciting with an ultrasound air-transducer (SMATR300H19XDA from Steiner \& Martins Inc). All measurements are subject to a systematic uncertainty of the interferometer of $\sim 5\,{\rm pm}$, and a statistical error determined by repeated measurements of $\sim 10\,{\rm pm}$, resulting in an error estimation on the displacements of $\sim 11.2\,{\rm pm}$. Careful error-propagation analysis results in error bars on all the figures which are smaller than the symbol size. The transducer has an essentially flat frequency response over the frequencies of interest, see Extended Data figure ~\ref{fig:ExtendedTransducer} (measured with a second air-transducer). The 0.46 dB variations are negligible with respect to the variations in response of 80 dB. 

To remove variations in response due to slight mis-alignments of the measurement point we normalize the local spectra by $\int d\nu\,\Delta z_i(\nu)\propto \int d\nu\,\psi_i^2(\nu)$, as required by the completeness of the eigenmodes. This is only valid under the assumption that all modes suffer from the same loss, or equivalently, have the same quality factor $Q\approx 1000$ (determined from the width of the corner modes). This assumption is justified for the following reason. Dissipation arises from two main sources: the viscoelasticity of the sample and the dissipation into the surrounding air. For both cases all disconnected plates suffer from the same damping. The perturbative nature of our beams (recall the bandwidth of $\sim 5 \, \rm{kHz}$ around the center frequency of $\sim 74\,{\rm kHz}$), restricts also the effects of the couplings on the dissipation. Moreover, our termination is such that all plates see an identical surrounding, independent of their location in bulk, along the edges or on the corners. Moreover, spectra based on data which is not normalized are almost identical to the ones shown in this paper (not shown). Finally, in all figures where arbitrary units are indicated, we normalize to the maximal value shown in the respective figure. 

As the bulk, edge, and corner modes spectrally overlap, there is no unique way to separate them in our measurements. However, the fact that the decay length is extremely short ($\xi/a\approx 1.6$, see above), a separation using the filters shown in Fig.~\ref{fig:midgap}d, where we simply select sites in the interior, along the edge and the corner sites respectively, is well justified.

\ifpreprint%
\else%
\begin{figure*}[tbh]
\includegraphics{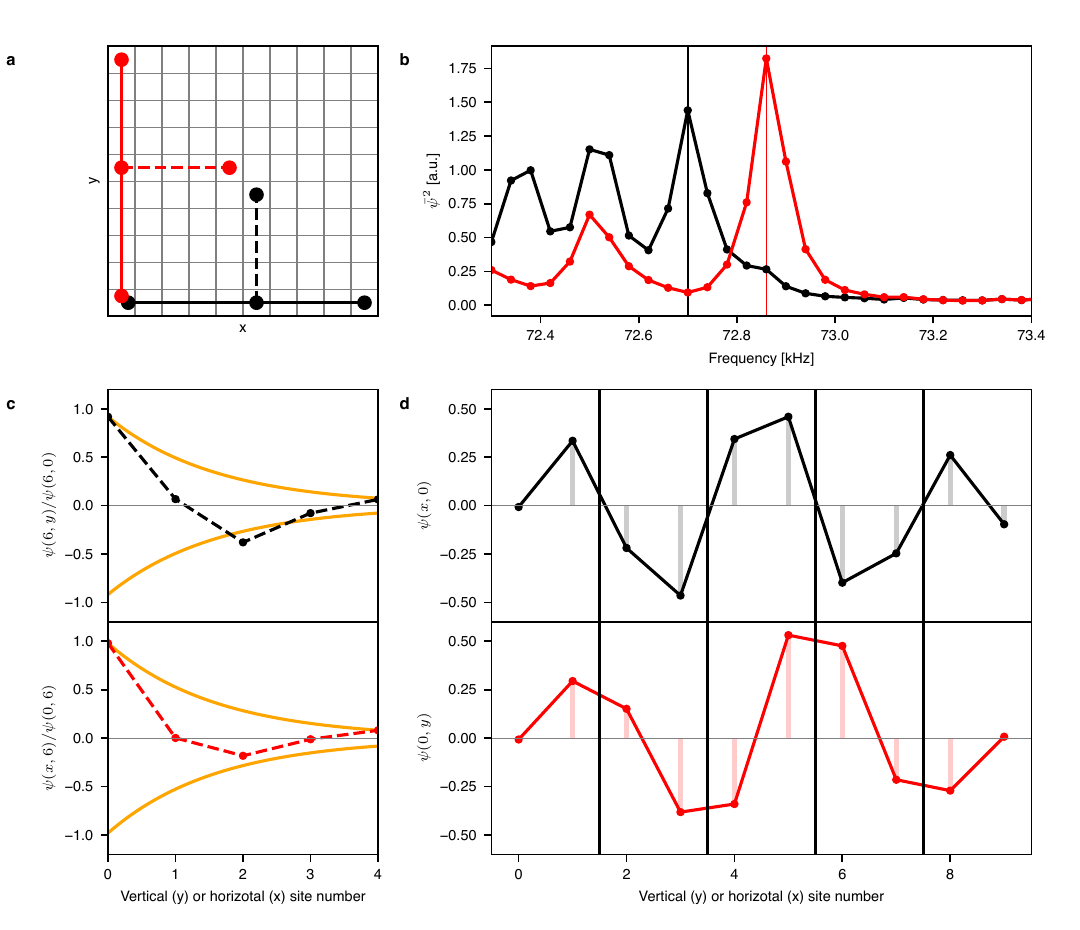}
\input{extendedfig4caption}
\label{fig:ExtendedGE}
\end{figure*}
\fi%
\noindent
{\bf Greens functions.}
In addition to the measurement of $\psi_i^2(\nu)$ by moving the exciter with the measurement point, we can also measure the Greens function $\psi_i(\nu)\psi_j(\nu)$ by fixing the exciter at site $j$ and moving the measurement point $i$, while exciting at frequency $\nu$. 

We first measure the corner Greens function for the four individual corners at their respective frequencies (determined from Fig.~\ref{fig:modes}d). In the Extended Data Fig.~\ref{fig:ExtendedG} we show the results. The density maps show the measured wave function $\psi(x,y)$ ($x$ and $y$ replace the site index $i$). The four panels proof that the four corner modes are independent and the spread in their frequency arises not from their hybridization. Along the edges we show the decay of the wave function and compare their envelope to the theoretical prediction with a decay length $\xi/a\approx 1.6$.

In the Extended Data Fig.~\ref{fig:ExtendedGE} we display the analysis of the edge physics by exciting on the bottom left corner and measuring along the lines indicated in the Extended Data Fig.~\ref{fig:ExtendedGE}a. The goal is to show that we can experimentally determine the sign of the couplings. To this end, we model our edge states by a simple Su-Schrieffer-Heeger model
\begin{linenomath}
\begin{equation}
    D(k) = 4\pi^2\nu_0^2+\sum_{i=1}^2 d_i(k) \sigma_i,
\end{equation}
\end{linenomath}
where the $\sigma$-matrices encode the two sublattices, $k$ is the momentum along the edge; $d_1(k)=\zeta[|\gamma|+|\lambda|\cos(k)]$ and $d_2=\zeta|\lambda|\sin(k)$. Along the horizontal edge, the couplings are positive $\zeta=1$, whereas along the vertical edge we have negative matrix elements $\zeta=-1$. The spectrum is given by $\omega_{\pm}(k)= \sqrt{4\pi^2\nu_0^2\pm \zeta|{\bf d}(k)|}$ with associated eigenvectors
\begin{linenomath}
\begin{equation}
        v_{\pm}(k) = \frac{e^{ikr_i}}{\sqrt{2}} \begin{pmatrix}\pm1 \\ \frac{ d_1(k)+ id_2(k)}{|{\bf d}(k)|} \end{pmatrix}.
\end{equation}
\end{linenomath}
Note that the highest frequency modes below the band gap are given by $\omega_-(\pi)$ for $\zeta>0$ and $\omega_+(\pi)$ for $\zeta<0$, respectively. For a finite edge, one can build eigenmodes from $v_{\pm}(k)$ that fulfill the desired boundary conditions. Note that the $\pm 1$ in the first component of $v_{\pm}(k)$ determines the relative sign between the mode inside one unit cell. Without specifying the exact boundary conditions, nor using knowledge on the values of $\gamma$ and $\lambda$, we cannot determine this relative sign. However, we can predict that it will be different on edges with $\zeta=\pm 1$.

To find the frequency of the highest mode per edge below the gap we show the integrated weight $\bar\psi^2=\sum_i \psi^2(\nu)$ along the respective edge in the Extended Data Fig.~\ref{fig:ExtendedGE}b. Fixing the excitation frequencies to the indicated values we measure the edge wave function these modes. The resulting sign change is indeed different (inside vs. between unit-cells) on the two edges. Finally, to further justify our filtering, we show that also the decay of the edge modes follows the expected decay with $\xi/a\approx 1.6$.

\ifpreprint%
\else%
\begin{figure*}[tbh]
\includegraphics{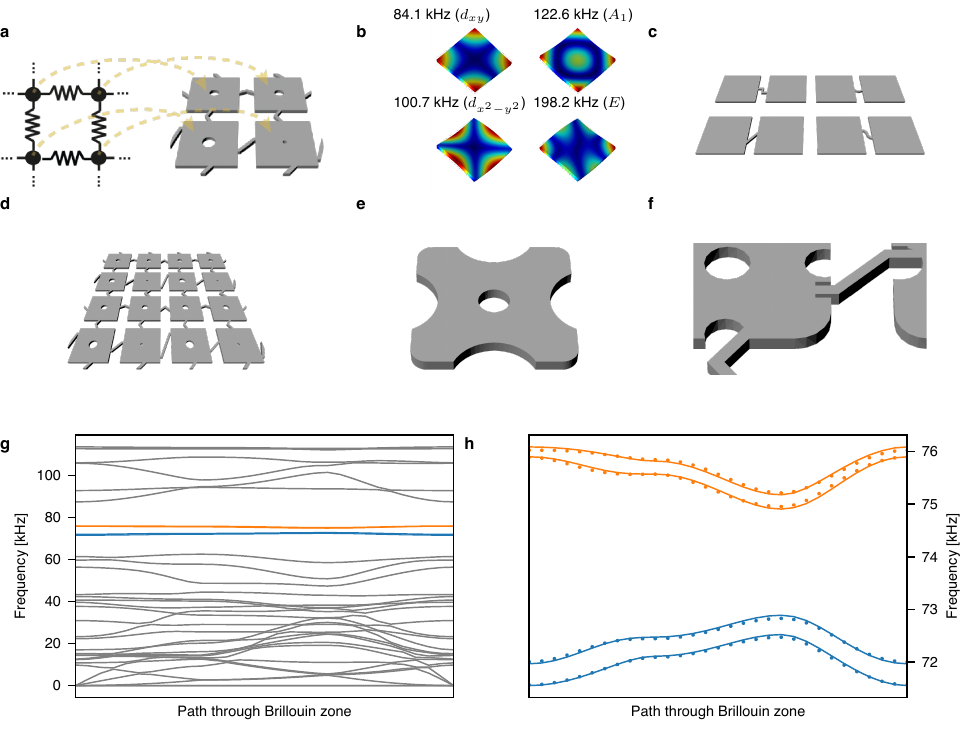}
\input{extendedfig5caption}
\label{fig:ExtendedDesign}
\end{figure*}
\fi%
\noindent
{\bf Sample design.} 
The plate geometries investigated in this article have been obtained in the framework of perturbative metamaterials, cf. Extended Data Fig.~\ref{fig:ExtendedDesign}.\cite{Matlack16} We combine geometric elements (silicon plates, beams and holes) to create a material that reproduces the discrete model of Benalcazar et al. over a range of frequencies. A perturbative metamaterial design consists of repeating basic resonating units (\paramPlateSize  silicon plates) that weakly interact with neighbouring resonant units. Here, this weak interaction is implemented using thin silicon beams. The weak interaction has two effects: First, the modes of isolated plates hybridize to Bloch bands of small bandwidth, preventing bands originating from unwanted modes to cross in frequency. Second, the weak interaction allows us to approximate the effect of different geometric elements by adding up individual contributions (See Ref. \onlinecite{Matlack16} for details), resulting in a drastic speedup of the calculation times.  

The design process starts by establishing a correspondence between the degrees of freedom in the metamaterial and those in the objective discrete model. This is done by expressing the dynamic deformation of the metamaterial's basic resonant units (plates) as a linear combination of free plate eigenmodes, see Extended Data Fig.~\ref{fig:ExtendedDesign}b. For sufficiently good spectral separation and sufficiently weak interactions, a single-mode local basis is enough to capture the material response with high precision. Each degree of freedom in the objective model is mapped to a single plate, which is assumed to vibrate in its first non-rigid-body, which for our parameters has $d_{xy}$ symmetry. Then, we evaluate individual coupling beam geometries to identify the most suitable designs and create a database relating beam geometry and coupling strength, obtained by simulating two-beam systems, cf. the Extended Data Fig.~\ref{fig:ExtendedDesign}c. Geometries are evaluated according to three parameters: (i) Ability to attain a broad range of couplings, (ii) low compressional strength to prevent the in-plane acoustic bands from reaching high frequencies where they  could hybridize with the topological band and (iii) absence of beam resonances in the frequency range of interest to exclude retardation effects in the couplings. Once the database has been assembled, we start a design by quickly constructing an approximate material geometry, and then refine it by performing a gradient optimization on an full model, cf. the Extended Data Fig.~\ref{fig:ExtendedDesign}d, that accounts for the interactions between different geometric features. 

We extract the effective theory for our design by first calculating the vibrational eigenmodes of a test system (Extended Data Fig. ~\ref{fig:ExtendedDesign}c) using the commercial Finite Element Method (FEM) package COMSOL Multiphysics. The eigenmodes' displacements along the three axes $u$, $v$ and $w$ are then interpolated over a regularly-spaced grid with a pitch of 0.05 mm. This interpolation is done for each mode $i$ and plate $j$, and denoted by $\psi_{ijk}$. A similar sampling is also performed for individual free-standing plates and denoted $\phi_{k}$ (here, the index $k$ labels the location and component of the displacement that is being interpolated). Once this information has been extracted from finite element simulations, the displacements of each degree of freedom for each mode are obtained by projecting the test system displacements into single-plate modes, $\alpha_{ij}=(\phi_{l}\phi_{l})^{-1}\phi_{k}\psi_{ijk}$ (repeated indices denote summation). After this procedure, the components of the matrix $\alpha_{ij}$ contain the displacements of the first non-rigid body mode of the $j$-th plate for the $i$-th eigenmode of the test system. The use of an interpolated grid allows us to use an individually optimized mesh for each finite element problem while still being able to express the results of one finite element simulation in terms of another's. 

The dynamic matrix $K$ describing the effective theory for the test system is obtained by $K_{lk}=\alpha_{ij}\Omega_{jl}^2\alpha_{lk}^{-1}$. Here, $\Omega_{jl}^2$ is a diagonal matrix whose elements contain the square angular frequencies of the modes in the frequency range of interest, $\Omega_{ii}^2=(2\pi f_{i})^2$. The resulting matrix $K$ has the same eigenfrequencies and projected eigenmodes as the full system and therefore provides a good description of the system's dynamics. This is highlighted in the Extended Data Fig.~\ref{fig:ExtendedDesign}g/h, which presents a comparison between the dispersion relation obtained from the effective theory and that obtained by solving a full finite-element model under Bloch boundary conditions.

\noindent
{\bf Sample fabrication.} 
The plate and beam geometry of Fig.~\ref{fig:setup}d implements the sought after weak and strong, positive and negative coupling matrix elements. The definition of $\gamma$ as the hopping strength {\em inside} a unit cell and $\lambda$ {\em between} unit cells renders $\gamma<\lambda$ the non-trivial phase. Connected to this identification is the notion of how we are allowed to terminate the system: Surfaces have to be compatible with the unit-cells, i.e., are not allowed to cut through unit-cells. In turn, this also means we can use the same design of Fig.~\ref{fig:setup}d and realize all phases shown in this paper by starting from a \site\!\!$\times$\site sample in the $(1/2,1/2)$-phase, then move the cut in $y$-direction by one row of sites to reach the $(1/2,0)$-phase. Finally we move the termination one column and end up in the $(0,0)$-phase. The coupling matrix elements are given by the ratio of the effective mass-density $\rho_{\rm\scriptscriptstyle eff}$ of the mode we use and the beam stiffness connecting two plates. We use a $\WthicknessMu\,\mu{\rm m}$ thick Si-wafer in (100) orientation, where we align the $x$- and $y$-axis of our model with the in-plane crystalline axes. The mass density of Si is $\rho=2330\,{\rm kg/m^3}$, the Young's moduli $E_x=E_y=E_z=130\,{\rm GPa}$, the Poisson ratios $\nu_{yz}=\nu_{zx}=\nu_{xy}=0.28$, and the shear moduli $G_{yz}=G_{zx}=G_{xy}=79.6\,{\rm GPa}$.\cite{Hall67} This results in an offset frequency for our mode of $\nu_0=\plateOffset\,{\rm kHz}$ and the coupling matrix elements are given by $\lambda=\strongHop\,({\rm rad}/{\rm s})^2$ and $\gamma=\weakHop\,\,({\rm rad}/{\rm s})^2$. The error estimation is detailed in the next paragraph.

Our samples are fabricated out of double side polished $100\,\rm{mm}$ Si-wafers. We measure individually the thickness of each wafer at several spots across the wafer, and we confirm that the overall total thickness variation within each wafer we use is $\leq 1\,\mu{\rm m}$. We fabricate plate and beam geometries as illustrated in Fig.~\ref{fig:setup}d using standard micro-fabrication techniques. First, $1\,\mu{\rm m}$ of SiO$_2$ is grown on the wafers via wet thermal oxidation (to be used as an etch mask), and a $2\,\mu{\rm m}$ thick layer of Al (that serves to protect the structure once the whole silicon has been removed) is deposited on the backside of the wafers using e-beam evaporation. A patterned $5\,\mu{\rm m}$ thick photoresist is used as an etch-mask when patterning the front side oxide in a reactive ion etching process. Using the remaining photoresist and the underlying oxide as etch masks, we etch through the wafer with a deep reactive ion etching following a Bosch$^{\text{\tiny\textregistered}}$ process alternating etching and passivation cycles. The ratio between both cycles is chosen to yield vertical side walls. This angle is characterized in several points of each wafer, confirming a variation of the angle of $\leq 2.5^\circ$. The Si etching terminates when reaching the backside oxide. The resulting oxide/aluminum membranes suspended between the beams and plates are removed by wet etching first the aluminum and then the oxide. This latter step also removes any oxide leftovers present on the front side. The main sources of of errors in the targeted model arising from the sample fabrication are: (i) Total thickness variation -- which we characterize being smaller than $1\,\mu{\rm m}$, and therefore stands for less than a 0.3\% variation across the wafer. (ii) Different sidewall angles between different parts of the wafer -- we measure it to be smaller than $2.5^\circ$. Hence, variation in feature sizes, when comparing front side to backside, may be up to $32\,\mu{\rm m}$. For the width of the plates this corresponds to an error of 0.3\%. Finally, (iii) the misalignment of the array with the material crystalline axis (100). This error has two sources, (a) wafer specifications indicate that the flap is located within $\pm 0.5^\circ$, and (ii) alignment during lithography, which the specifications of our machine states around $\pm 1^\circ$. In either case, this results in an overall error of less than 0.1\% in the Young’s modulus. This error leads to the stated uncertainties in the local plate frequencies and couplings using standard elasticity theory. Finally, the wafers are clamped between two steel plates (each of $3\,{\rm mm}$ thickness), cf. Fig.~\ref{fig:midgap}c. The impedance miss-match between the steel plates and the wafer leads essentially to fixed boundary conditions $\Delta z=0$.

\end{document}